\documentclass[twocolumn]{aastex63}

\newcounter{daggerfootnote}

\newcommand{\muv}{$M_{\mathrm {UV}}$}
\newcommand{\jwst}{\textit{JWST}}
\newcommand{\hst}{\textit{HST}}

\definecolor{mycol}{rgb}{0,0,1}

\received{n/a}
\revised{n/a}
\accepted{n/a}

\submitjournal{\apjl}

\shorttitle{GLIMPSE Discovery of $z>16$ galaxies}
\shortauthors{Kokorev et al.}

\usepackage{xcolor}
\usepackage{longtable}
\usepackage{lipsum}
\usepackage{amsmath}
\usepackage[normalem]{ulem}
\usepackage[para]{threeparttable}
\usepackage{enumitem}
\usepackage[encapsulated]{CJK}

\begin{document}

\title{A Glimpse of the New Redshift Frontier Through Abell S1063}

\correspondingauthor{Vasily Kokorev}
\email{vkokorev@utexas.edu}

\author[0000-0002-5588-9156]{Vasily Kokorev}
\affiliation{Department of Astronomy, The University of Texas at Austin, Austin, TX 78712, USA}

\author[0000-0002-7570-0824]{Hakim Atek}
\affiliation{Institut d'Astrophysique de Paris, CNRS, Sorbonne Universit\'e, 98bis Boulevard Arago, 75014, Paris, France}

\author[0000-0002-0302-2577]{John Chisholm}
\affiliation{Department of Astronomy, The University of Texas at Austin, Austin, TX 78712, USA}

\author[0000-0003-4564-2771]{Ryan Endsley}
\affiliation{Department of Astronomy, The University of Texas at Austin, Austin, TX 78712, USA}

\author[0009-0009-9795-6167]{Iryna Chemerynska}
\affiliation{Institut d'Astrophysique de Paris, CNRS, Sorbonne Universit\'e, 98bis Boulevard Arago, 75014, Paris, France}

\author[0000-0002-8984-0465]{Julian B.~Mu\~noz}
\affiliation{Department of Astronomy, The University of Texas at Austin, Austin, TX 78712, USA}

\author[0000-0001-6278-032X]{Lukas J. Furtak}
\affiliation{Department of Physics, Ben-Gurion University of the Negev, P.O. Box 653, Be'er-Sheva 84105, Israel}

\author[0000-0002-9651-5716]{Richard Pan}
\affiliation{Department of Physics \& Astronomy, Tufts University, MA 02155, USA}

\author[0000-0002-4153-053X]{Danielle Berg}
\affiliation{Department of Astronomy, The University of Texas at Austin, Austin, TX 78712, USA}

\author[0000-0001-7201-5066]{Seiji Fujimoto}
\affiliation{Department of Astronomy, The University of Texas at Austin, Austin, TX 78712, USA}
\affiliation{Cosmic Dawn Center (DAWN), Niels Bohr Institute, University of Copenhagen, Jagtvej 128, K{\o}benhavn N, DK-2200, Denmark}

\author[0000-0001-5851-6649]{Pascal A. Oesch}
\affiliation{D\'epartement d'Astronomie, Universit\'e de Gen\`eve, Chemin Pegasi 51, 1290 Versoix, Switzerland}
\affiliation{Cosmic Dawn Center (DAWN), Niels Bohr Institute, University of Copenhagen, Jagtvej 128, K{\o}benhavn N, DK-2200, Denmark}

\author[0000-0001-8928-4465]{Andrea Weibel}
\affiliation{D\'epartement d'Astronomie, Universit\'e de Gen\`eve, Chemin Pegasi 51, 1290 Versoix, Switzerland}


\author[0000-0002-8192-8091]{Angela Adamo}
\affiliation{Department of Astronomy, The Oskar Klein Centre, Stockholm University, AlbaNova, SE-10691 Stockholm, Sweden}

\author[0000-0003-1609-7911]{Jeremy Blaizot}
\affiliation{Universite Claude Bernard Lyon 1, CRAL UMR5574, ENS de Lyon, CNRS, Villeurbanne, F-69622, France}

\author[0000-0002-4989-2471]{Rychard Bouwens}
\affiliation{Leiden Observatory, Leiden University, NL-2300 RA Leiden, Netherlands}

\author[0000-0003-0348-2917]{Miroslava Dessauges-Zavadsky}
\affiliation{D\'epartement d'Astronomie, Universit\'e de Gen\`eve, Chemin Pegasi 51, 1290 Versoix, Switzerland}

\author[0000-0002-3475-7648]{Gourav Khullar}
\affiliation{Department of Physics \& Astronomy and PITT PACC, University of Pittsburgh, Pittsburgh, PA 15260, USA}

\author[0000-0002-3897-6856]{Damien Korber}
\affiliation{D\'epartement d'Astronomie, Universit\'e de Gen\`eve, Chemin Pegasi 51, 1290 Versoix, Switzerland}

\author[0009-0007-8470-5946]{Ilias Goovaerts}
\affiliation{Space Telescope Science Institute, 3700 San Martin Dr., Baltimore, MD 21218, USA}

\author[0009-0004-4725-8559]{Michelle Jecmen}
\affiliation{Department of Astronomy, The University of Texas at Austin, Austin, TX 78712, USA}

\author[0000-0002-2057-5376]{Ivo Labb\'e}
\affiliation{Centre for Astrophysics and Supercomputing, Swinburne University of Technology, Melbourne, VIC 3122, Australia}

\author[0000-0002-6085-5073]{Floriane Leclercq}
\affiliation{Univ Lyon, Ens de Lyon, CNRS, Centre de Recherche Astrophysique de Lyon UMR5574, F-69230, Saint-Genis-Laval, France}

\author[0000-0001-8442-1846]{Rui Marques-Chaves}
\affiliation{D\'epartement d'Astronomie, Universit\'e de Gen\`eve, Chemin Pegasi 51, 1290 Versoix, Switzerland}

\author[0000-0002-3407-1785]{Charlotte Mason}
\affiliation{Cosmic Dawn Center (DAWN), Niels Bohr Institute, University of Copenhagen, Jagtvej 128, K{\o}benhavn N, DK-2200, Denmark}

\author[0000-0001-5538-2614]{Kristen B.~W.\ McQuinn}
\affiliation{Space Telescope Science Institute, 3700 San Martin Dr., Baltimore, MD 21218, USA}
\affiliation{Department of Physics \& Astronomy, Rutgers, The State University of New Jersey, Piscataway, NJ 08854, USA}

\author[0000-0003-3729-1684]{Rohan Naidu}
\affiliation{MIT Kavli Institute for Astrophysics and Space Research, 70 Vassar Street, Cambridge, MA 02139, USA}

\author[0000-0002-5554-8896]{Priyamvada Natarajan}
\affiliation{Department of Astronomy, Yale University, 219 Prospect Street, New Haven, CT 06511, USA}
\affiliation{Black Hole Initiative at Harvard University, 20 Garden Street, Cambridge, MA 02138, USA}

\author[0000-0002-7524-374X]{Erica Nelson}
\affiliation{Department for Astrophysical and Planetary Science, University of Colorado, Boulder, CO 80309, USA}

\author[0000-0002-7534-8314]{Joki Rosdahl}
\affiliation{Universite Claude Bernard Lyon 1, CRAL UMR5574, ENS de Lyon, CNRS, Villeurbanne, F-69622, France}

\author[0000-0001-8419-3062]{Alberto Saldana-Lopez}
\affiliation{Department of Astronomy, The Oskar Klein Centre, Stockholm University, AlbaNova, SE-10691 Stockholm, Sweden}

\author[0000-0001-7144-7182]{Daniel Schaerer}
\affiliation{D\'epartement d'Astronomie, Universit\'e de Gen\`eve, Chemin Pegasi 51, 1290 Versoix, Switzerland}

\author[0000-0002-6849-5375]{Maxime Trebitsch}
\affiliation{LERMA, Sorbonne Université, Observatoire de Paris, PSL Research University, CNRS, 75014 Paris, France}

\author[0000-0002-3216-1322]{Marta Volonteri}
\affiliation{Institut d'Astrophysique de Paris, CNRS, Sorbonne Universit\'e, 98bis Boulevard Arago, 75014, Paris, France}

\author[0000-0002-0350-4488]{Adi Zitrin}
\affiliation{Department of Physics, Ben-Gurion University of the Negev, P.O. Box 653, Be'er-Sheva 84105, Israel}

\begin{abstract}
We report the discovery of two galaxy candidates at redshifts between  $15.7<z<16.4$ in \jwst\ observations from the GLIMPSE survey. These robust sources were identified using a combination of Lyman-break selection and photometric redshift estimates. The ultra-deep NIRCam imaging from GLIMPSE, combined with the strong gravitational lensing of the Abell S1063 galaxy cluster, allows us to probe an intrinsically  fainter population (down to \muv $=-17.0$ mag) than previously achievable. These galaxies have absolute magnitudes ranging from \muv $= -17.0$ to $-17.2$ mag, with blue ($\beta \simeq -2.87$) UV continuum slopes, consistent with young, dust-free stellar populations. The number density of these objects, log$_{\rm 10}$($\phi$/[Mpc$^{-3}$ mag$^{-1}$])=$-3.47^{+0.13}_{-0.10}$ at $M_{\rm UV}=-17$ is in clear tension with pre-\jwst\ theoretical predictions, extending the over-abundance of galaxies from $z\sim10$ to  $z\sim 17$. These results, together with the scarcity of brighter galaxies in other public surveys, suggest a steep decline in the bright-end of the UV luminosity function at $z \sim 16$, implying efficient star formation and possibly a close connection to the halo mass function at these redshifts. Testing a variety of star formation histories suggests that these sources are plausible progenitors of the unusually UV-bright galaxies that \textit{JWST} now routinely uncovers at $z = 10-14$. Overall, our results indicate that the luminosity distribution of the earliest star-forming galaxies could be shifting towards fainter luminosities, implying that future surveys of cosmic dawn will need to explore this faint luminosity regime.
\end{abstract}


\section{Introduction} \label{sec:intro}
According to the standard paradigm of structure formation, the same primordial fluctuations that gave rise to hot and cold spots in the cosmic microwave background (CMB) will eventually grow, collapse, and form the first galaxies during cosmic dawn, ushering in the epoch of first light \citep[e.g.][]{loeb13}. These first galaxies have remained outside of our observational reach for decades. That is because they are faint and highly redshifted. Even the deepest \textit{Hubble Space Telescope} ({\hst}) surveys have fallen short of observing first light \citep{oesch16}. The ultraviolet (UV) light from the first galaxies drops precipitously due to absorption by foreground neutral gas (the Lyman-$\alpha$ break, see e.g.,~\citealt{steidel96}), which makes galaxies above $z\gtrsim 10$ invisible in the \hst\, IR filters.

The \textit{James Webb Space Telescope} (\textit{JWST}) was designed to observe the first galaxies. 
With its enormous collecting area and unprecedented near infrared imaging and spectroscopic capabilities, \textit{JWST}/NIRCam is sensitive to faint light up to 5~$\mu$m \citep{rieke23}. These are exactly the wavelengths required to discover galaxies forming stars during the first few hundred million years of cosmic history, as their Lyman-$\alpha$ break falls at 2.0~$\mu$m and 2.5~$\mu$m at redshifts 16 and 20, respectively.
This means that broad imaging bands, such as the F200W and F277W can be paired to search for continuum ``dropouts'' up to $z\sim 20$, a mere 180 million years after the Big Bang.

\textit{JWST} has successfully used this technique to discover dozens of galaxies at $10<z<15$. These observations have revealed a surprising over-abundance of UV-bright galaxies at $z > 10$, challenging pre-JWST predictions \citep[to name just a few,][]{adams23,austin23,atek23a,curtislake23,donnan23,finkelstein23, perezgonzalez23,castellano24, carniani24,mcleod24, robertson24, chemerynska24}.
Attempts to explain this over-abundance of bright galaxies at extreme redshifts invoke extremely efficient star formation in the early Universe, bursty star formation histories, a lack of dust attenuation, a top-heavy initial mass function (IMF), more efficient formation of dark matter halos, or modifications to the $\Lambda$CDM paradigm \citep[e.g.,][]{pacucci22,boylan-kolchin23,dekel23, ferrara23, finkelstein23,  harikane23_agn,sun23,li24_ffb}.

The ancestors of these UV-bright galaxies at even higher redshift ($z>15$) have remained elusive. Many observational campaigns have searched for detections of first light galaxies, but few candidates pass rigorous scrutiny. Extremely dusty star-forming galaxies \citep{naidu22_fix,zavala23,arrabal_haro23}, and spurious detections can masquerade as faint F200W drop-out galaxies. A key issue with detecting first-light galaxies is that they are expected to be intrinsically faint as they have had little time to assemble significant stellar mass. The discovery and characterization of these elusive primordial galaxies provides empirical constraints on the astrophysics shaping their formation.

Here we report on observations from the \textit{JWST} GLIMPSE survey (Atek et al., 2025, in prep.). GLIMPSE was specifically aimed to detect galaxies during the epoch of first light, by pairing ultra-deep NIRCam imaging across seven wide and two medium bands with the gravitational lensing of the foreground galaxy cluster Abell S1063. With even modest magnification factors of 2, the 30.8 mag GLIMPSE observations can reach absolute magnitudes ($M_{\rm UV}$) of $\sim -17$ at $z\sim16$, probing the faint galaxies that likely existed during the epoch of first light. We use these observations to identify two robust $z\gtrsim16$ candidates, observed within a single NIRCam pointing (\autoref{fig:fig_field}), and assess the validity of these candidates via model fitting, morphology and properties of their stellar populations.

\par
Throughout this work we assume a flat $\Lambda$CDM cosmology with $\Omega_{\mathrm{m},0}=0.3$, $\Omega_{\mathrm{\Lambda},0}=0.7$ and H$_0=70$ km s$^{-1}$ Mpc$^{-1}$, and a \citet{chabrier} initial mass function (IMF) between $0.1-100$ $M_{\odot}$. All magnitudes are expressed in the AB system \citep{oke74}.

\section{Observations and Data} \label{sec:obs_data}
Detailed descriptions of the observations, data reduction, cluster light removal, and source extraction will be presented in the GLIMPSE survey (Atek et al., 2025, in prep.); and are briefly summarized below.

\subsection{GLIMPSE Survey}
This work uses the ultra-deep imaging from the public GLIMPSE survey (PID: 3293, PIs: H. Atek \& J. Chisholm). GLIMPSE targets Abell S1063 (AS1063 hereafer), one of the highest-magnification regions in the Hubble Frontier Fields \citep[HFF;][]{lotz17}, with 7 broadband filters (F090W, F115W, F150W, F200W, F277W, F356W, F444W) and 2 medium-band filters (F410M, F480M). The total duration of the survey is $\sim155$h of science time, reaching unprecedented observed depths, down to 30.6 mag uniformly across all wide bands. Specifically, the bands used in the $z>15$ drop-out selection - F200W and F277W have integration times of 19 and 23 hours respectively. Furthermore, the lensing magnification of Abell S1063 also means that we can probe intrinsically faint sources which would otherwise be invisible in the deepest \textit{JWST} blank field surveys.

GLIMPSE observations use a MEDIUM8 readout pattern for all exposures, and adopt a 6 position primary dither pattern to cover the short wavelength intra-module gaps while maximizing the full-depth area. In addition, we use a subpixel dither with 4 positions to best sample the PSF. This was done to optimize the S/N, with dithers large enough to mitigate the fixed pattern noise, imperative when searching for high-$z$ targets.

\subsection{Data Reduction}
\label{sec:data_red}
NIRCam imaging data for all 7 broad and 2 medium band filters in AS1063 are reduced following the procedure in \citet{endsley24} using the \texttt{jwst\_1293.pmap} context map. We implement crucial enhancements over the standard STScI pipeline, including corrections for cosmic rays, stray light, 1/f noise, and detector artifacts \citep{bradley23, rigby23}. Furthermore, we construct our own set of image flats based on all public NIRCam imaging as of January 12, 2025. Given that GLIMPSE adopts a sub-pixel dithering pattern to minimize overheads, any impurities in the flats can be co-added in the dithering procedure and thus might appear as real sources in each primary dither position. Compared to the publicly available flats provided by the Space Telescope Science Institute available as of mid-January 2025 (first released in September 2023), our flats result in substantially improved depth in the final mosaics. We recover an $\approx0.3-0.5$ mag improvement in depth across all long-wavelength bands, and an $\approx0.1-0.2$ mag improvement across the short-wavelength bands. Finally, given the depth of the GLIMPSE campaign and the presence of bright cluster galaxies (bCGs), we model and subtract the background on an amplifier basis with \textsc{sep} \citep{sep}, while manually masking out the bright regions. This further improves the image background and allows the mosaics to reach $\sim0.3$ mag deeper across all filters.

Our final data-set achieves $5\sigma$ aperture-corrected nominal depths of 30.8--30.9 mag across all broad-bands in D=0\farcs{2} apertures.

We also process and incorporate the deep \textit{HST} ACS and WFC3 mosaics from the Hubble Frontier Fields \citep{lotz17} and BUFFALO \citep{steinhardt20} programs. The \textit{HST} images are based on Gaia-aligned mosaics from the CHArGE archive \citep{kokorev22}, which are hosted on the Dawn JWST Archive \citep{valentino23}. We drizzle our final mosaics onto a 0\farcs{02}/pixel grid for the \textit{JWST} short wavelength (SW) filters, and  0\farcs{04}/pixel for \textit{JWST} long wavelength (LW) and \textit{HST}.

\begin{figure*}
\begin{center}
\includegraphics[width=.89\textwidth]{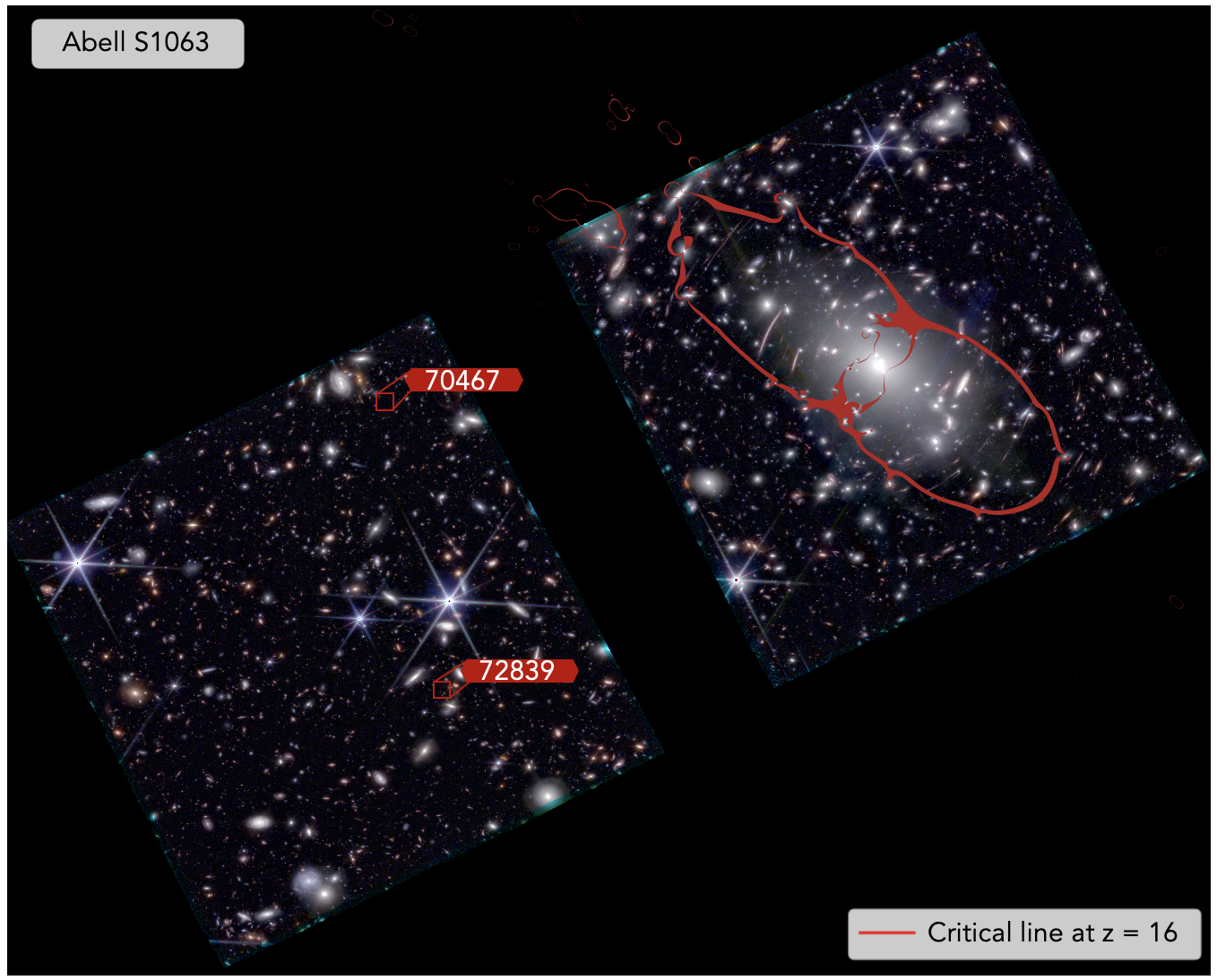}
\caption{\textbf{GLIMPSE field overview.} Positions of the high-$z$ candidates are overlaid on top of the RGB image constructed from all available \textit{JWST} broad bands. All of the high-$z$ candidates are located in full coverage (6 primary dithers) area of Module A (left). We additionally overlay a $z=16$ critical curve in red.}
\label{fig:fig_field}
\end{center}
\end{figure*}

\subsection{Cluster Light Contamination}
Careful handling of the contamination light from the bCGs and intra-cluster light (ICL) is necessary for our science objectives. Improper treatment of the ICL could potentially lead to inaccurate colors which in return affect the derived photometric redshifts, negatively impacting the high-$z$ selection in the proximity of the cluster. To model and subtract the bCG and ICL light, we follow the robust methods described in \cite{ferrarese06}, \cite{shipley18} and \cite{weaver24} for the Hubble Frontier Fields-Deep Space (HFFDS) and UNCOVER/Mega Science photometric catalogs \citep{suess24}. After both the bCGs and the ICL have been modeled and subtracted, we perform an additional local background subtraction pass in the affected areas, largely following the same methodology as in \autoref{sec:data_red}. 

\subsection{Source Extraction}

We construct empirical point-spread functions (PSFs) by stacking the available stars in the field, and match all available bCG-subtracted \textit{JWST} and \textit{HST} images to our lowest resolution PSF - F480M. The sources are detected using \textsc{SExtractor} \citep{sextractor}, which we run in a single mode on an inverse-variance weighted combination of the native PSF F277W, F356W and F444W images. We then use \textsc{photutils} \citep{photutils} to measure flux densities in circular apertures with varying diameters from $D=0\farcs{1}$ -- $1\farcs{2}$. The photometric errors are determined separately for each object and filter, taking into account both the aperture size and the depth variation across the image. In the vicinity of each source, we place 2000 random apertures in source-free parts of the image (as determined by the segmentation map).
The standard deviation of the flux density within empty apertures, plus the Poisson noise are then used as the final uncertainty. This method is generally preferable over just using the weight/error maps, as it better accounts for effects of correlated noise \citep[e.g. see][]{endsley23,weaver24}. The aperture corrections are calculated by assuming a point-source profile. We use the empirical F480M symmetric PSF curve of growth to determine the fraction of the total flux that falls outside of each specified aperture size and then use that as our correction factor.

In order to select high-fidelity high-$z$ candidates, we flag all sources that fall on the edge of the mosaic, intersect with diffraction spikes from bright stars, or are close ($<2\farcs{0}$) to the modeled and subtracted bCGs. As galaxies at high-$z$ are generally small, we will only use the total fluxes within a $D=0\farcs{2}$ aperture for the remainder of our work.

\begin{figure}
\begin{center}
\includegraphics[width=.45\textwidth]{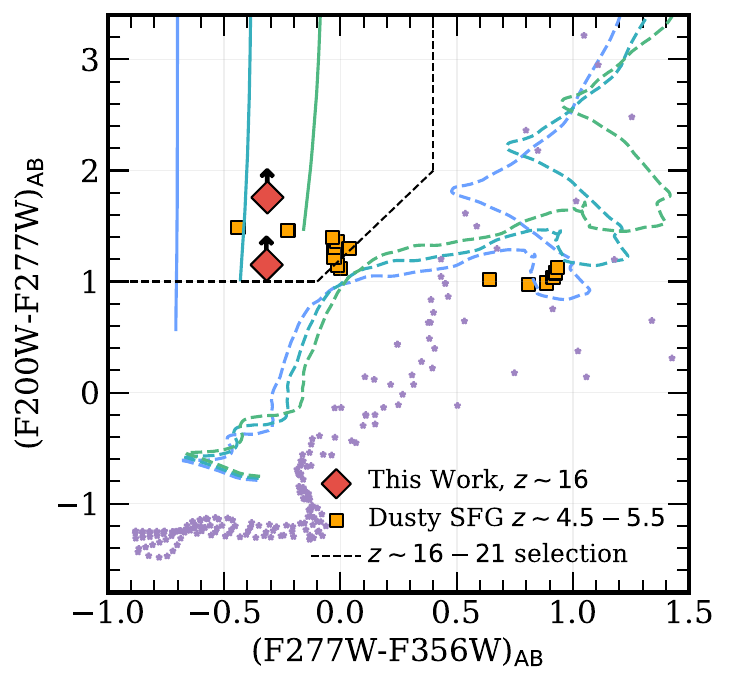}
\caption{\textbf{High-$z$ color-color selection.}  Our final high-$z$ sample, which we also present in \autoref{fig:fig_sample} is shown with red diamonds. The color-color Lyman break selection is indicated with black dashed line. Solid lines show the tracks followed by starburst galaxies at $z\sim16$ \citep{chevallard16} with varying levels of attenuation $A_{\rm V}$ = 0--0.5 (shown with blue to green). We highlight low-redshift quiescent galaxies \citep{polletta07}, whose Balmer break could mimic the Lyman break at higher redshifts, as dashed lines. Extreme dusty starbursts \citep{naidu22_fix,arrabal_haro23,zavala23} at $z\sim4.5-5.5$ (orange squares) look quite similar to a high-$z$ galaxy with this selection alone (see \autoref{sec:lowz} for why these are disfavored). Finally, cool stars and brown dwarfs \citep{chabrier00,allard01} are shown in purple.}
\label{fig:fig_color}
\end{center}
\end{figure}

\section{Data Analysis} \label{sec:data_analysis}
Our goal is to find galaxies above $z > 16$. Previous experience has demonstrated that a single selection technique, however, is insufficient to provide robust $z > 16$ candidates, and might result in low-$z$ contaminants \citep[e.g. see][]{naidu22_fix}. To mitigate that, we combine our selection to use the Lyman Break (\autoref{sec:break}) and photometric redshift test (\autoref{sec:photoz}) which stringently tests for low-z contamination (\autoref{sec:lowz}) and uses the size estimates (\autoref{sec:size}) to further test their robustness.

\subsection{Lyman Break Selection}
\label{sec:break}
The first selection applied to the photometric catalog is based on identifying dropouts in the rest-frame UV, using the Lyman-$\alpha$ break technique. We define a selection window in color-color space to isolate dropouts between the F277W and F200W filters, while simultaneously excluding objects with red continua in the rest-frame UV between the F277W and F356W filters. These color criteria are determined from running synthetic photometry on a set of galaxy templates generated using {\sc BEAGLE} \citep[][]{chevallard16}. We used starburst templates in the redshift range $15<z<20$, incorporating IGM attenuation following \citet{inoue14}, with varying levels of attenuation between $A_{V}=0$ and $A_{V}=0.5$, based on the SMC extinction law \citep{gordon03}. We also consider low-redshift quiescent galaxies that could mimic a Lyman break, using SEDs from the SWIRE template library \citep{polletta07}. In addition, we include cool stars and brown dwarfs, using stellar templates from \citet{chabrier00} and \citet{allard01}. The color-color tracks for all these simulated sources are shown in \autoref{fig:fig_color}, along with the adopted selection window, which is defined by the following criteria:

 \begin{equation*}
	\begin{array}{l}
		M_{200}-M_{277}>1.0\\
		M_{200}-M_{277}>1.2+2.0(M_{277}-M_{356})\\
		M_{277}-M_{356}<0.5
	\end{array}
\end{equation*}

In addition to the color criteria, we require that sources are well detected (at the $>3\sigma$ level in all three of the broad-band LW filters), and have a SNR of $\gtrsim4$ in at least one band, while remaining undetected (at the $<2\sigma$ level) in bands blueward of the Lyman break (F090W, F115W, and F150W). This first selection leads to a sample of $\sim38$ $z\gtrsim16$ galaxy candidates.

\subsection{Photometric Redshift}
\label{sec:photoz}
To calculate photometric redshifts ($z_{\rm phot}$) for all objects in the GLIMPSE catalog, we use the \textsc{Python} version of \textsc{EAZY} \citep{brammer08}. We choose the 
\textsc{blue\_sfhz\_13} model subset\footnote{\url{https://github.com/gbrammer/eazy-photoz/tree/master/templates/sfhz}} that contains redshift-dependent SFHs, and dust attenuation values. More specifically, the linear combinations of log-normal SFHs included in the template set are not allowed to exceed redshifts that start earlier than the age of the Universe \citep[for more detail see][]{blanton07}. These models are further complemented by a blue galaxy template, derived from a \textit{JWST} spectrum of a $z=8.50$ galaxy with extreme line equivalent widths \citep{carnall22}.

Our work is concerned only with the F200W dropouts, for which the \textit{HST} data are simply too shallow to be useful.  Therefore, in our analysis we only focus on the GLIMPSE \textit{JWST} bands. We fit the aperture corrected $D=0\farcs{2}$ \textit{JWST} flux densities, including the upper limits, using the $0.01<z<30$ redshift grid. The uncertainties on the photometric redshifts are computed from the 16th and 84th percentiles of the redshift probability distributions - $p(z)$. The best-fit \textsc{EAZY} SEDs are only used to validate the color-color high-$z$ selection.

In order to select $z>16$ galaxy candidates with \textsc{EAZY} we then require that the sources must be detected (S/N$>3$) in at least 3 bands, similarly to our color-color selection, and have a well constrained $p(z)$ (FWHM$<2.5$) without a statistically significant secondary redshift solution. Cross-referencing our \textsc{EAZY} high-$z$ sample with the color-color selected F200W dropout from \autoref{sec:break} results in 8 candidates at $z\simeq 15.6 - 19.5$.

In addition to \textsc{EAZY}, we use \textsc{BEAGLE} \citep{chevallard16} and the methods outlined in \citet{endsley24}, to independently derive photometric redshifts. Briefly, we use the \textsc{BEAGLE} models constructed from updated \citet{bruzual03} stellar templates that use the \texttt{PARSEC} isochrones \citep{bressan12, chen15}. These models have been passed through \texttt{Cloudy} \citep{ferland17} to self-consistently produce nebular (both line and continuum) emission \citep{gutkin16}. We explore different parametric star formation histories (including constant, burst, and a two-component star formation history), stellar ages (between 1~Myr and 30~Gyr),  metallicities (0.0063--0.5~Z$_\odot$), and attenuation laws \citep[we settle on the SMC law;][]{pei92}. While the two codes agree exceptionally well (\autoref{fig:fig_sample}) for the majority of our high-$z$ candidates, we further remove 2 sources where the \textsc{BEAGLE} fit shows a prominent secondary peak solution, leaving us with 6 candidates.

\subsection{Low-$z$ Confusion}
\label{sec:lowz}
The early days of \textit{JWST}, initial discoveries saw a slew of $z>16$ galaxy candidates, for which valid concerns have been raised in the literature regarding the ability of photometric redshifts to pick out true high-$z$ candidates \citep{arrabal_haro23,donnan23,zavala23}. These works have shown that dusty extremely star-forming galaxies at $z\sim5$ can masquerade as high-$z$ objects and bias our view of high-redshift galaxies.  For example, the so-called ``Schr\"{o}dinger'' galaxy was initially photometrically-identified at $z>16$ \citep{naidu22_fix,donnan23}, and then had its redshift revised to a $z\sim5$, via extreme emission lines found in NIRSpec spectra, as an interloper \citep{arrabal_haro23,zavala23}. Despite this misidentification, these outliers provide invaluable insight into the potential pitfalls of our methodology.
Subsequently, identification methodologies have been improved, and the techniques used to process (e.g. various pipeline improvements and calibrations) and analyze \textit{JWST} data have significantly matured, such that the vast majority of $\sim$ 95 photometrically-identified $z>10$ galaxies have been spectroscopically confirmed at high-redshift \citep[][]{arrabal_haro23,hainline24,harikane24}. At this moment, these initial misidentifications provide a road map to rigorously scrutinize the $z>16$ candidates, however at the moment no galaxies with $z>15$ have been confirmed spectroscopically.

To address the potential contamination from low-$z$ interlopers, we re-fit all of our 9 high-$z$ candidates with the \textsc{EAZY} \textsc{blue\_sfhz\_13} model suite, plus an additional dusty starburst template \citep{naidu22_fix,arrabal_haro23}, to mimic the low-$z$ interlopers. The redshift grid in this case is forced to be $z<6$, to match the redshift range of dusty interlopers. Deciding which model is better relies on the Bayesian Information Criterion (BIC) test \citep{schwarz78}. This test computes which template is the most likely fit to the observed data, while penalizing models that have too many free parameters. In our case a high-$z$ fit is statistically preferred over forced low-$z$ ($z<6$) solutions with the additional ``Schr\"{o}dinger'' template, when a BIC difference reaches 3 or more \citep[using the criteria defined in][]{jeffreys61}. We find that a majority -- 5/6 -- of our high-$z$ objects can be fit equally well with a low-$z$ dusty starburst template ($\Delta$BIC$<3$, corresponding to $\gtrsim3\sigma$ significance). We are now left with 5 $z\sim16$ galaxy candidates. While these appear to be robustly detected ($\sim3\sigma$) in 3 bands, we further limit our final sample to only include galaxy candidates that reach at least $5\sigma$ in one or more bands. Doing this further removes 3 objects that are located on the edges of our mosaics, where we do not have the full coverage of all 6 primary dithers (see \autoref{fig:fig_field}), and thus do not achieve the full depth. We further verify that our remaining objects are not simply hot pixels in the LW bands, by separately examining all 6 primary dithers in the F277W band, as well as the LW stack.

This final quality cut ensures that only the best high-$z$ candidates end up in our final sample. This leaves us with 2 objects which we show in \autoref{fig:fig_sample}. It is worth pointing out that the secondary $z<6$ solution with a ``Schr\"{o}dinger'' template is choosing exactly the redshift range ($z\sim 4.9$) where emission lines can confuse the redshift fitting codes \citep{naidu22_fix,arrabal_haro23}, however given our chosen $\Delta$BIC threshold, high-$z$ \textsc{EAZY} solutions are still preferred. To further test this, we carry out an additional \textsc{BEAGLE} low-$z$ fit where we very finely sample the $z=4.8-5.1$ range in an attempt to find the highly specific redshift solution where high emission lines intensities can mimic a break. Our results find that the the finer spacing grid does not change the conclusion in a statistically significant way. Finally we note that the morphology of our objects (\autoref{fig:fig_sample}), does not change from band-to-band when moving to redder filters, as it would for dusty galaxies.

\subsection{Transient Contamination}

\begin{figure*}
\begin{center}
\includegraphics[width=.72\textwidth]{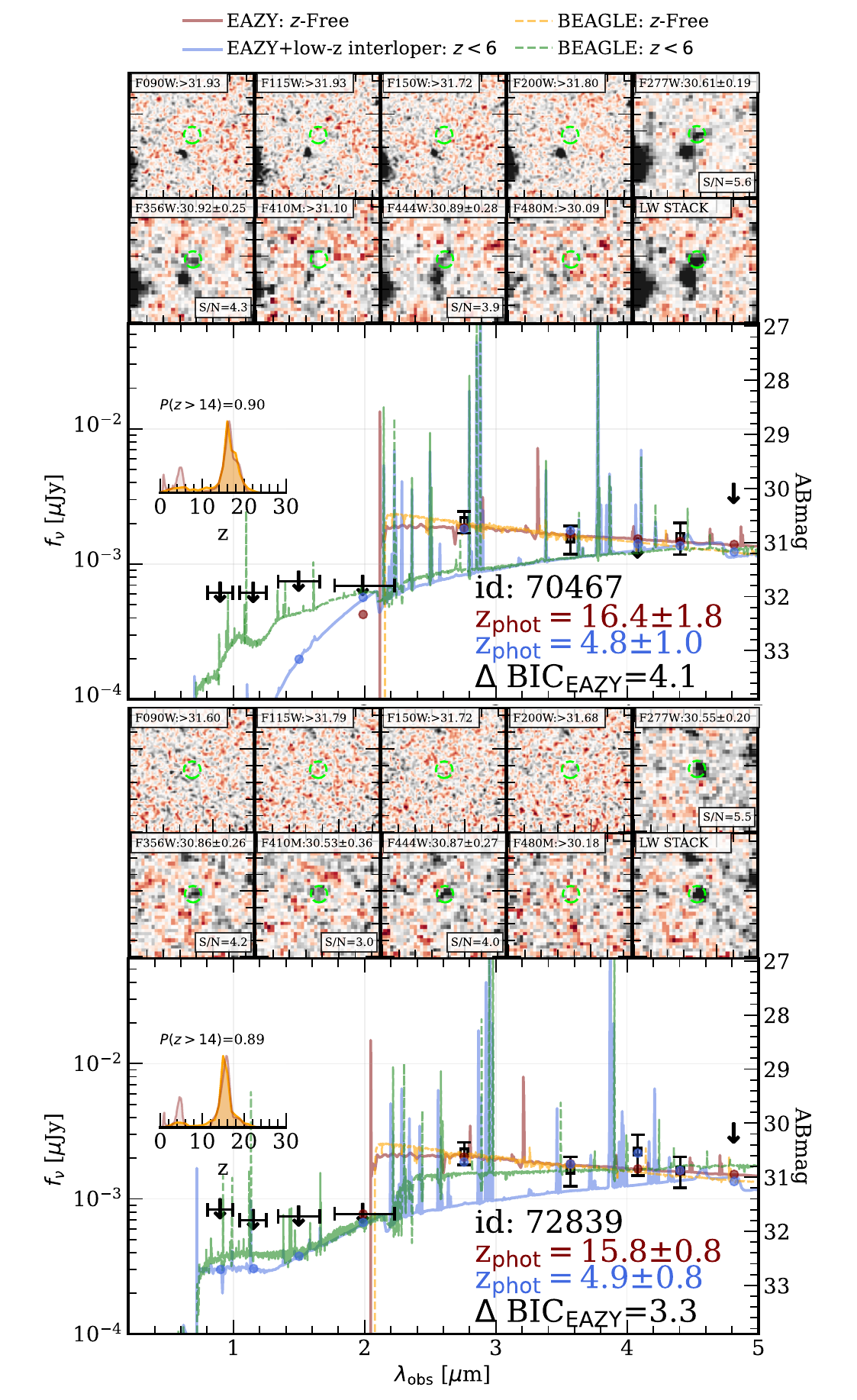}
\caption{\textbf{Most secure $z\gtrsim16$ candidates from GLIMPSE.} For each source we show 1\farcs{5} cutouts in all broad and medium band NIRCam filters, plus a detection (F277W+F356W+F444W) LW stack. On each cutout we overlay the extraction aperture (D=0\farcs{2}) in green. On the SED plot, we show the best fit high-$z$ vs low-$z$ \textsc{EAZY} models (maroon and blue), as well as the same for \textsc{BEAGLE} (dashed orange and green). The colored circles represent the integrated flux density for each solution. Detections are shown as black squares, $3\sigma$ upper limits are shown as downward arrows. We explicitly highlight the difference between a low-$z$ interloper and high-$z$ EAZY solution with a $\Delta$BIC statistic on each panel, as discussed in \autoref{sec:lowz}. In the inset panel we show the $p(z)$ for both \textsc{EAZY} and \textsc{BEAGLE} fits with a free redshift solution. Finally, we display the probability of these candidates being at $z>14$ as derived from multiplying \textsc{EAZY} and \textsc{BEAGLE} redshift distribution.}
\label{fig:fig_sample}
\end{center}
\end{figure*}

Low-$z$ confusion could also arise from transient events, as was noted in \citet{decoursey25} who showed that certain supernovae can mimic SEDs of $z>16$ galaxies. Generally, multi-epoch observations with long enough baselines can rule out these contaminants, however GLIMPSE data with only $\sim24$ hours between LW observations are not suitable to detect such changes. Despite that, as \citet{decoursey25} points out, deeper observations in SW channels can aid in distinguishing a supernova from a Lyman break, as the drop off of the former is more gradual. Given the sharp breaks, aided by deep F200W data, and extended morphologies of our candidates (\autoref{fig:fig_sample}), contamination by supernovae seems unlikely.

\subsection{Gravitational Magnification}
\label{sec:mag}
To take into account the effects from gravitational lensing we use a new strong lensing (SL) model of AS1063 constructed for GLIMPSE with the updated version of the \citet{zitrin15} parametric method that was recently used for several \textit{JWST} SL clusters \citep[e.g.][]{pascale22,furtak23,meena23}. We model the cluster with two smooth dark matter (DM) halos parametrized as pseudo-isothermal elliptical mass distributions \citep[PIEMDs;][]{kassiola93}: one centered on the bCG, and the other on a group of galaxies in the north-east of the cluster \citep[e.g.][]{bergamini19,beauchesne24}. In addition, we model 303 cluster member galaxies as dual pseudo-isothermal ellipsoids \citep[][]{eliasdottir07}. The model is constrained with 80 multiple images of 30 sources, 26 of which have spectroscopic redshifts \citep[][]{balestra13,monna14,richard21,beauchesne24,topping24}. The model achieves an average image reproduction error of $\Delta_{\mathrm{RMS}}=0.47\arcsec$. We refer the reader to Furtak et al. (in prep.) for more details on the lens model. A previous (\textit{HST}-based) version of this model was recently used in \citet{topping24}.

Magnifications are computed analytically at the position of each galaxy candidate and adopting its photometric redshift. The magnifications and their uncertainties are listed in \autoref{tab:tab1}.

\subsection{Size Measurements}
\label{sec:size}

The cutouts we show in \autoref{fig:fig_sample} imply that our sources are quite compact, yet appear to be resolved. To measure the effective radii we model each source with \textsc{GALFIT} \citep{peng02,peng10}, using a S\'ersic \citep{sersic} profile where the source position, brightness, effective radius, S\'ersic index, and axis ratio are allowed to vary. When fitting, we take into account the effects of the PSF, which we measure empirically from the bright stars in the field. We perform this procedure on our brightest band - F277W, to ensure optimal S/N per pixel is achieved to accommodate robust size measurements. Additionally, in the case of source 70464, we simultaneously model both the high-$z$ object and the low-$z$ ($z_{\rm phot}\sim3$) neighboring galaxy. We find that the on-the-sky sizes of our sources, range from $R_{\rm eff}\sim0\farcs{08}-0\farcs{10}$. A source can be considered to be resolved when its effective radius is larger than the empirical PSF half-width at half maximum (HWHM). Since the HWHM of F277W PSF is $\sim$0\farcs{046}, we can consider all our sources to be resolved. 
 
Taking the redshift and gravitational magnification into account, we convert our angular sizes to physical effective radii. These are roughly similar for both objects with $R_{\rm eff} \sim 200$ pc. We list the de-lensed sizes in \autoref{tab:tab1}.

\section{Results} \label{sec:res}
\subsection{Stellar Population Properties}
After completing the multiple stages of selecting our final $z>16$ sample, we now compute a range of relevant physical parameters. The uniqueness of our sample lies in the unprecedented depth of the GLIMPSE observations, which we further push to their limit with gravitational lensing in order to obtain these candidates. On average, our sources have 3 individual band detections, with the rest of the \textit{JWST} photometry being upper limits. While this is perfectly adequate to constrain the redshift, deriving any physical parameters, especially stellar mass, from SED fitting codes would be simply unreasonable. As such, we will only focus on the  
observable stellar population parameters that can be derived from the rest-UV photometry alone.

\begin{figure}
\begin{center}
\includegraphics[width=.47\textwidth]{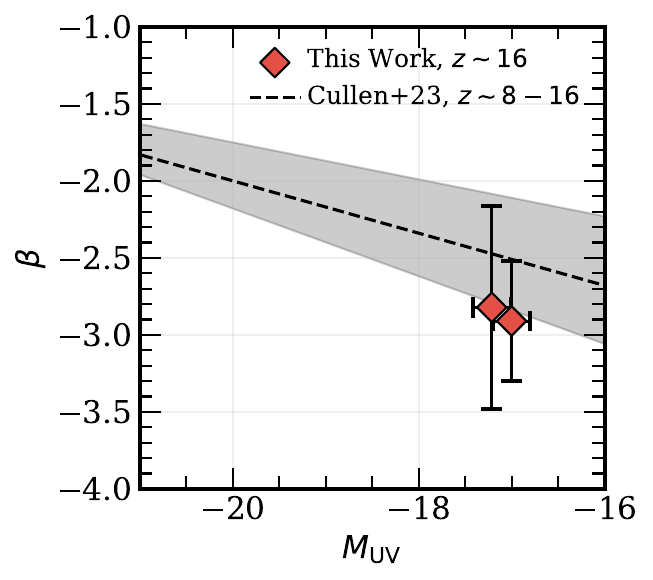}
\caption{\textbf{UV continuum slope and de-lensed luminosity.} The black dashed line shows the best fitting $\beta$-$M_{\rm UV}$ relation from \citet{cullen23} for SFGs at $z=8-16$. Grey shaded area corresponds to the 68 \% confidence interval. Our secure ($>5\sigma$) sources at $z\sim16$, shown with red diamonds, appear to be consistent with the predicted trend within 1-sigma, although systematically shifted to bluer slopes.}
\label{fig:beta}
\end{center}
\end{figure}

We estimate the absolute UV magnitude for each galaxy from the observed F277W band, which samples $\lambda_{\rm rest}\sim1500$ \AA\, at this redshift. We find that our targets cover a very narrow range of derived $M_{\rm UV}=-17.1^{+0.10}_{-0.12}$, after accounting for lensing magnification. We derive the UV-slope $\beta$ for each object by assuming $f_{\lambda}\sim\lambda^{\beta}$, and fitting it to all the photometric points that fall within the $\lambda_{\rm rest}$ in 1260 -- 2500 \AA\, range, which effectively traces the observed F277W-F356W color. We find that our $\beta$ values span a very narrow range from $-2.9$ to $-2.8$, with a median of $-2.87\pm0.15$. These are shown in \autoref{fig:beta}, alongside the magnification corrected UV brightness.
Finally, we derive the SFR$_{\rm UV}$ directly from our de-lensed $M_{\rm UV}$, by following the relation from \citep{kennicutt12}. We find that the SFR$_{\rm UV}$ ranges from 0.4 to 0.6 $M_{\odot}$ yr$^{-1}$. Since our derived $\beta$ values imply negligible dust reddening, we do not apply a correction for dust when deriving the SFR. All the de-lensed physical parameters are listed in \autoref{tab:tab1}.

\begin{deluxetable}{ccc}[]
\tabcolsep=2mm
\tablecaption{\label{tab:tab1} Properties of the high-$z$ sample$^\dagger$. }.
\tablehead{Parameter & 70467 & 72839}
\startdata
$\rm{RA [deg]}$ & 342.2386 & 342.2319 \\
$\rm{Dec [deg]}$ &  -44.5347 & -44.5565 \\
$z_{\rm phot}$ & $16.4\pm1.8$ & $15.8\pm0.8$\\
$\Delta \chi^{2}$ & 1.0 & 1.2 \\
$\Delta$BIC$^1$ & 4.1 & 3.3 \\
$\mu$ & $1.55\pm0.04$ & $1.35\pm0.02$ \\
$M_{\rm UV}$ [ABmag] & $-17.0\pm0.2$ & $-17.2\pm0.2$\\
SFR [M$_\odot$/yr] & $0.4\pm0.1$ & $0.5\pm0.1$ \\
$\beta$ & $-2.91\pm0.20$ & $-2.82\pm0.33$ \\
$R_{\rm eff}$ [pc] & $212\pm23$ & $199\pm71$ \\
\enddata
\begin{tablenotes}
\footnotesize{$^\dagger$  All values are corrected for gravitational magnification.} \\
\footnotesize{$^1$ BIC(low-z) - BIC(high-z).} \\
\end{tablenotes}
\end{deluxetable}

\subsection{Number Density}
\label{sec:nd}
We compute the number density  of $16<z<20$ sources in GLIMPSE by utilizing a 1/$V_{\rm max}$ method \citep{schmidt}, where $V_{\rm max}$ corresponds to the maximum volume a galaxy could occupy and still be detected in the appropriate filter for our redshift range. Due to lensing, the effective area, and therefore volume, covered by our observations, is smaller than it would be in a blank field. In addition, an accurate derivation of the UV number density requires a robust estimate of the completeness of our survey, for a given selection function and depth. To account for the former, we use our magnification maps together with the $p(z)$ for our sources (to account for the redshift range) and including the uncertainties on both derive a total survey volume of $17055\pm71$ Mpc$^{3}$.

The second step to derive the effective volume consists of computing the survey completeness through the lensing cluster. To do that, we use the same approach adopted in \citet{atek18} and \citet{chemerynska24}. Briefly, the procedure includes generating a large set of mock galaxies that are distributed directly in the source plane, which in turn was generated from our lensing model. The properties of the simulated objects span the redshift, size and luminosity range of our objects. More details regarding this procedure will be presented in Chemerynska et al. in prep. Combining this with our derived volume, we find an effective survey volume $V_{\rm max}=5882.7\pm1538.9$ Mpc$^{3}$. The final uncertainty on the volume is derived from the combined errors on the lensing model (to be presented in Furtak et al. in prep.), the redshift probability distribution of the high-$z$ candidates and completeness simulation itself.

All of our objects cover nearly the same range of $M_{\rm UV}$, so to derive the final number density we will just assume a single bin with a width of 1 magnitude -- $M_{\rm UV}=-17.0\pm0.5$. The final number density for our sample in the $16<z<20$ range is therefore equal to log$_{\rm 10}$ ($\phi$/[Mpc$^{-3}$ mag$^{-1}$])=$-3.47^{+0.13}_{-0.10}$.

\begin{figure}
\begin{center}
\includegraphics[width=.45\textwidth]{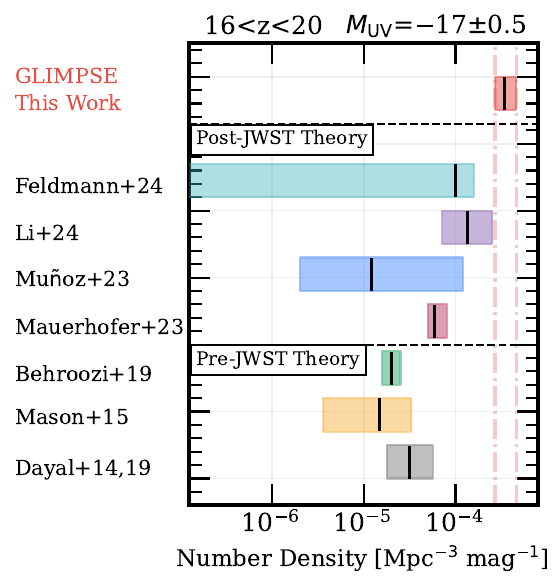}
\caption{\textbf{Theoretical comoving number densities of $z>16$ galaxies in the literature.} Our completeness-corrected densities are shown in red. Black lines are median values, shaded areas show 68 \% percentiles. Values have been homogenized in terms of redshift interval ($16<z<20$) and de-lensed. While high, our predictions generally align with the upper edge of the theoretical predictions \citep{mason15,dayal14,dayal19,behroozi19,mauerhofer23,munoz23,feldmann24,li24_ffb}.}
\label{fig:nd_theory}
\end{center}
\end{figure}

\begin{figure}
\begin{center}
\includegraphics[width=.48\textwidth]{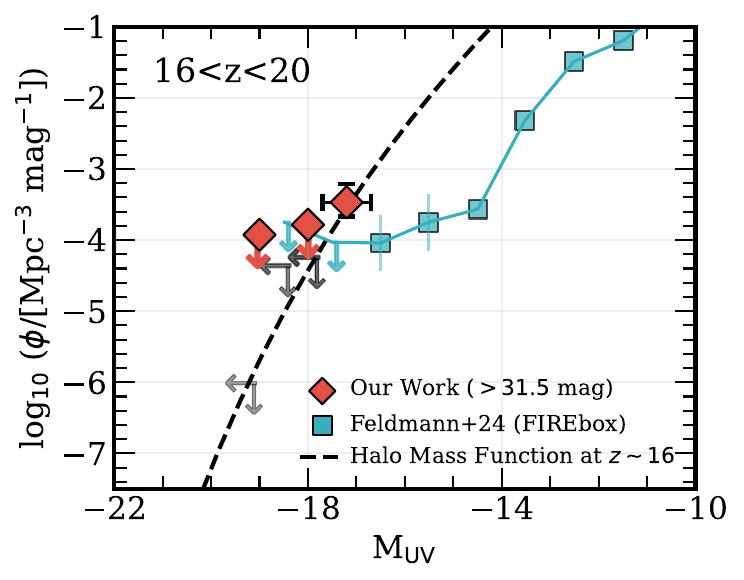}
\caption{\textbf{Observed abundance of bright galaxies at $z>16$.} We report the approximate magnitude limits for a number of deep extragalactic surveys corresponding to the $2\sigma$ depth of the dropout filter - F200W, which is required for the Lyman break identification. In varying shades of gray, we show CEERS \citep{bagley23,finkelstein23} and PRIMER \citep[both COSMOS and UDS][]{donnan23} combined, JADES (Origins Field, \citealt{eisenstein23}) and NGDEEP \citep{leung23,bagley24}. The dashed line is a scaled version of the Halo Mass Function that assumes a 30\% star formation efficiency of the gas, no dust, and a continuous star formation history. The dashed line is  simply a toy model to illustrate how the UVLF at these redshifts could  steeply decline.}
\label{fig:uvlf}
\end{center}
\end{figure}

\begin{figure}
\begin{center}
\includegraphics[width=.49\textwidth]{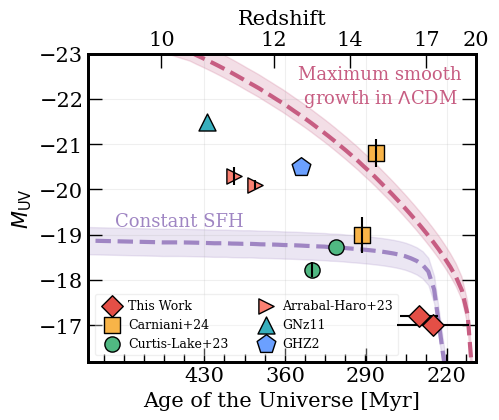}
\caption{\textbf{The abundance of bright sources at cosmic dawn.} We trace the possible evolutionary paths of our objects from $z\sim16$, while factoring in the redshift uncertainty, to some of the brightest spectroscopically confirmed high-$z$ objects at $z\simeq11$--14 \citep{arrabal_haro23,bunker23,curtislake23,carniani24,castellano24}. Shown are the tracks (with $1\sigma$ uncertainty represented by shaded regions) of constant star-formation at $\sim1$ $M_{\odot}$/yr (purple), and a $\Lambda$CDM-limited maximal possible accretion scenario \citep[pink line,][]{dekel13}, where all gas is converted to stars. The overabundance of bright galaxies observed at high-$z$ so far is fully consistent with our observations. We only show the $M_{\rm UV}$ uncertainties that are larger than the marker size.}
\label{fig:sfh}
\end{center}
\end{figure}

\section{Discussion} \label{sec:disc}
\subsection{Stellar Populations at High Redshift}
In this section, we reflect on the implications of our derived parameters if the selected sample truly resides at $z\sim16$. 

First, we compare our calculated $R_{\rm eff}$ with empirical results for spectroscopically confirmed galaxies at $z\sim13-14$. Both \citet{carniani24} and \citet{curtislake23} report UV sizes in the range of $\sim100-300$ pc, consistent with our findings. While high-redshift size predictions from simulations are limited, the TNG50 simulations have accurately reproduced galaxy morphologies across a wide redshift range \citep[e.g., see][]{tacchella19}. The latest high-$z$ size evolution analyses in \citet{constantin23} and \citet{morishita23} suggests a scaling relation of $\sim8.66\times(1+z)^{-1.15}$ kpc. Extrapolating this to $z\sim16$ predicts $R_{\rm eff}\sim 250-300$ pc, aligning with our measurements within $1\sigma$. Generally, the relatively large sizes of our objects may indicate that UV light from these galaxies arises from extended stellar populations, as observed in \citet{carniani24}. Furthermore, the spatially resolved nature of our sources, combined with an absence of a PSF-like, centrally concentrated component, suggests that the UV emission is not dominated by unobscured AGN, unlike other high-$z$ galaxies \citep{maiolino23,harikane23_agn}.

Using our derived SFR$_{\rm UV}$ and $R_{\rm eff}$, we calculate the UV-based star formation rate surface density, finding a median value of $\Sigma_{\rm SFR, UV}=1.14^{+0.20}_{-0.18}$ M$_{\odot}$ yr$^{-1}$ kpc$^{-2}$. This is $5-10$ times lower than values reported for spectroscopically confirmed objects at $z\sim14$ by \citet{carniani24} and is similarly lower when extrapolating the trend in \citet{calabro24}. However, higher $\Sigma_{\rm SFR, UV}$ found at $z<15$ likely reflect observational biases and small number statistics at high-$z$, with the bulk of the population likely containing less intense star-formation. Moreover, if the accretion rate changes exponentially with redshift for galaxies such as ours, even small changes in redshift will likely lead to significantly more accretion. We will explore this further in \autoref{sec:bright_gal}.

The UV slopes in our sample also do not exhibit unusual characteristics. We find a median $\beta$ of $-2.87$, consistent with a largely dust-free stellar plus nebular continuum. This is reasonable, as our objects are observed just $\sim200$ Myr post-Big Bang, with limited time for substantial dust production. Within the uncertainties of our $\beta$ values, our sources align with the $\beta$-$M_{\rm UV}$ luminosity relation from the surveys used in \citet{cullen23}, which are based on recent \textit{JWST} data for $z=8-16$ star-forming galaxies. Finally, we do not observe extremely negative beta slopes that have been postulated for metal-free Population III stars \citep[e.g. see][]{jaacks18}.

\subsection{Where are the Bright High-$z$ Galaxies?}
The GLIMPSE imaging data are contained within a single NIRCam pointing, with the effective area further limited by lensing effects. Despite the relatively small volume surveyed, we identify 2 robust high-$z$ candidates, all exhibiting characteristics consistent with a star-forming population at this epoch. We now discuss the number density derived from our sample, comparing it with theoretical predictions from both pre- and post-\textit{JWST} launch models (see \autoref{fig:nd_theory}).

Regardless of the simulation type—whether semi-analytic \citep{dayal14,mason15,behroozi19} or hydrodynamical \citep{rosdahl22,wilkins23}—we find that our derived number density at $16<z<20$ at $\mathrm{M_{\rm UV}=-17.0\pm0.5}$ significantly exceeds the pre-JWST theoretical expectations, with a discrepancy of over $3-5\sigma$. This mirrors an observed discrepancy in the UVLF suggested by spectroscopically confirmed $z\sim12-14$ objects \citep{donnan23, finkelstein23, casey23,carniani24, chemerynska24, harikane24, robertson24}. In contrast, post-\textit{JWST} calibrated simulations, including semi-analytic models \citep{mauerhofer23,munoz23}, analytic predictions that include the feedback-free burst scenario \citep{li24_ffb}, and FIREbox hydrodynamical simulations \citep{feldmann24}, show better, generally within $1-2\sigma$, agreement.

We find the best agreement with the FIREbox simulations presented in \citet{feldmann24}. This might suggest that the abundance of UV-luminous galaxies at $z\gtrsim11$ \citep{arrabal_haro23,maiolino23,curtislake23,castellano24,carniani24} could be due to a relatively constant star-formation efficiency (SFE) that is largely independent of $M_{\rm halo}$ (I. Chemerynska in prep.). We will explore the connection of our sample to these lower-$z$ UV-bright galaxies in the next section. Encouragingly, our sample, while showing somewhat higher number densities at this absolute magnitude, remains broadly consistent with many post-\textit{JWST} theoretical predictions.

Over the past two years, \textit{JWST} has identified dozens of bright $M_{\rm UV}<-18$ high-$z$ galaxy candidates at $z=10-14$, challenging pre-\textit{JWST} expectations \citep[][]{adams23,austin23,atek23a,curtislake23,donnan23,finkelstein23, castellano24, carniani24, robertson24, chemerynska24,mcleod24,leung23,perezgonzalez23}. Surprisingly, despite the large areas and depths of various surveys, none have detected bright galaxies beyond $z>16$. Given the detection limits of $\sim30-30.5$ mag in the F200W/F277W filters (used for Lyman break dropout selection at $z>15$), bright galaxies with $M_{\rm UV}<-19$ would likely have been identified by these surveys if they indeed existed at these redshifts. 
Selecting these galaxies at $z>16$ requires deep F200W (drop-out band) and F277W photometry, as the break has to be unambiguous ($>1$ mag). To compute the expected number densities from various extragalactic surveys, based on non-detections \citep{gehrels86} we do the following. Using the quoted F200W $2\sigma$ depths for CEERS \citep{bagley23,finkelstein23}, PRIMER (both COSMOS and UDS), JADES Origins Field \citep{eisenstein23, robertson24}, and NGDEEP \citep{leung23,bagley24}, we assume M$_{200} - $M$_{277} \geq 1$~mag and then compute the expected  $M_{\rm UV}$ and the corresponding number density (from $V_{\rm max}$) for each survey using the reported survey area. These are shown in \autoref{fig:uvlf} as gray upper limits, alongside the GLIMPSE number density at $M_{\rm UV}=-17$ and upper limits at $M_{\rm UV}=-19$ and $-18$. The discrepancy between the limits at $M_{\rm UV}<-18$ and our observed number densities at $M_{\rm UV}=-17$ is quite jarring. The upper limits reach $\sim 10^{-6}$ Mpc$^{-3}$ mag$^{-1}$ (for CEERS plus PRIMER) at $M_{\rm UV}\sim-19$, in a stark contrast to GLIMPSE where we predict there are $\gtrsim2000\times$ more galaxies per volume at $M_{\rm UV}=-17$. If $M_{\rm UV}\lesssim-18$ galaxies existed at this epoch, they surely should have been detected by now.

So where are they? One possibility is that because of cosmic variance, the current deep fields have missed some of the bright $z > 15$ galaxies. After all, the depths of GLIMPSE are comparable to JADES and NGDEEP, as such we would expect these surveys to find a $z >15$ source. In this scenario, wider-field surveys that reach M$_{\rm UV} \sim -20$~mag would discover bright $z>15$ galaxies. For instance, objects brighter than $M_{\rm UV}\sim-18$ are too rare to appear in the FIREbox simulated volume at this redshift (see \autoref{fig:uvlf}). 

Furthermore, this difference can be potentially due to the bright end of the UVLF at $z>16$ lying exactly at the GLIMPSE detection limit of $M_{\rm UV}\sim-17$, explaining why we are seeing these objects for the first time (though see a candidate in NGDEEP found by both \citet{leung23} and \citet{austin23} at $z_{\rm phot}\sim 15.6$, with $M_{\rm UV}$ of $\sim-19.2$, which implies a log$_{\rm 10}$ ($\phi$/[Mpc$^{-3}$ mag$^{-1}$])$\sim-4.65$). We would also like to further highlight several relatively bright ($M_{\rm UV} \sim -19$) $z > 16$ photometric candidates identified in \citet{hainline24} and \citet{conselice24}, which suggest comparable number densities to those found in \citet{leung23} and \citet{austin23}. However, these objects exhibit weaker Lyman breaks than the candidates in our study. Before any of these objects are confirmed spectroscopically, the fate of the UVLF at high-$z$ would remain elusive.

The steepness of the UVLF implied by the GLIMPSE detections against the brighter limits can be fit if the UVLF closely tracks the halo mass function, as we show in \autoref{fig:uvlf} (translating halo mass to SFR with a toy model given by the cosmic baryon fraction, zero dust, and 30\%  efficiency of gas into stars over a fixed 100 Myr timescale). We note that the most plausible way to reproduce the observed number density is with a constant and relatively high SFE.

The lensing magnification present in AS1063, as opposed to deep blank fields such as JADES and NGDEEP, could also play a role in this discrepancy. We speculate that even the modest 
lensing magnification of $\mu\sim1.5$, which will double the exposure time on source (adding $\sim0.5$ mag in depth), will provide the additional sensitivity needed to uncover these early galaxies. There simply was no feasible way to see these relatively rare objects prior to GLIMPSE.

\subsection{Abundance of Bright Galaxies at $z\sim12-14$}
\label{sec:bright_gal}
Our observations indicate that one possible explanation for the lack of $M_{\rm UV}\lesssim-18$ galaxies at $z\sim16$ is that brighter galaxies are extremely rare at high-$z$, so only ``faint'' galaxies would be abundant enough to be found in the volume probed so far at this epoch. Therefore one can imagine that the over-abundance of bright galaxies observed at $z=12-14$ have likely evolved from earlier (fainter) galaxies, such as the ones presented here, that are surprisingly abundant for their relatively faint UV luminosities. In return, this suggests that the discrepancy between theoretical predictions and observations may begin as early as $z=18$, potentially during the epoch when the first galaxies formed.

To test that, we would like to see how different SFHs starting at $z\sim16$ reproduce the observed bright galaxies at $z<14$. To do this we consider two cases. First, we consider a constant SFH which 
begins forming stars at the median redshift of our sample ($z\sim16.1$) at a rate of $\sim 1$ $M_{\odot}$ yr$^{-1}$ (to match our derived SFRs). As the second case, we would consider the maximum growth rate allowable for these galaxies by forming stars at 100\% of the accretion rate set by the Extended Press-Schechter approximation in a $\Lambda$CDM Universe \citep[e.g., see eq. 7 in ][with a 0.3 dex scatter as in \citealt{ren18} and \citealt{Mirocha:2020slz}]{dekel13}. This in turn creates an exponentially increasing star formation history that puts a maximum limit on the luminosity growth experienced by these galaxies observed  at these redshifts. Since we want to show the absolute limit possible, given our observations, we start to evolve this SFH at the 84th percentile ($1\sigma$) of the median redshift of our sample - $z\sim18.0$, instead of simply the median. Both scenarios begin at $M_{\rm UV}\sim-17$. To evolve these SFHs we have used the toolkit within \textsc{BAGPIPES} \citep{carnall19}, which outputs $M_{\rm UV}$ in a pre-defined grid of lookback-time. We show the results of our modeling, alongside the bright spectroscopically confirmed galaxies in \autoref{fig:sfh}.

We find that if our galaxies followed a constant SFH and continued forming stars at their current rate, they can easily reproduce the objects observed in \citet{curtislake23} and the fainter one of the two galaxies presented in \citet{carniani24}. Reproducing some of the brightest ($M_{\rm UV}<-20$) known galaxies at $z\geq10$ would, however, require very (nearly 100\%) efficient star-formation episodes. Following the maximal exponential accretion scenario, we find that galaxies in our sample can, within $1\sigma$, become as bright as JADES-GS-z14-0 ($M_{\rm UV}=-20.81$ at 
$z=14.17933$; \citealt{carniani24,carniani24b,schouws24}), or GNz11 \citep[although the stellar contribution to the $M_{\rm UV}$ could be lower, due to the potential AGN;][]{oesch16,bunker23,maiolino23}, without a need to invoke any exotic cosmological framework. Curiously, the SFH for JADES-GS-z14-0 presented in \citeauthor{carniani24} predicts SFR $\lesssim 1$ $M_\odot$ yr$^{-1}$ at $z\sim16-17$, which matches our sample if it were a potential progenitor. This further reemphasizes that these $z\sim16$ candidates could plausibly grow to become the extremely bright $z\sim14$ galaxies without extraordinarily new cosmologies.

\section{Conclusions}
In this study, we leverage ultra-deep GLIMPSE NIRCam imaging covering the AS1063 lensing cluster to identify a unique sample of high-redshift ($z > 16$) galaxy candidates. Using a combination of observed colors alongside an extensive suite of SED modeling routines, we select two robust candidates at $z\simeq15.8-16.4$, which all fit into our single NIRCam pointing. 

We examine the stellar population properties of our sample, such as their delensed, absolute UV-brightness, $\beta$ slopes and sizes, finding that these high-$z$ galaxies fit very well into the picture of galaxy evolution being established at $z\sim10-14$. Crucially, our findings reveal a significant excess in number density of galaxies at $M_{\rm UV}=-17$ of log$_{\rm 10}$ ($\phi$/[Mpc$^{-3}$ mag$^{-1}$])=$-3.47^{+0.13}_{-0.10}$, compared to theoretical predictions from both semi-analytic and hydrodynamical simulations calibrated before the \textit{JWST} launch, mirroring similar discussions for bright galaxies at lower $z\sim10-14$ \citep[e.g. see][]{oesch16,carniani24,finkelstein23,arrabal_haro23}.

Surprisingly, none of the current extragalactic \textit{JWST} surveys have managed to securely identify brighter objects beyond $z > 16$. No bright ($M_{\rm UV}\lesssim-18$) galaxy candidates have been identified by substantially larger areas covered by CEERS and PRIMER \citep{bagley23,finkelstein23,donnan23}. The same is the case for moderately large and deep observations from JADES \citep{eisenstein23} and NGDEEP \citep{leung23,bagley24}. What our observations seem to indicate is that this dearth of bright galaxies at high-$z$ is driven by the shape of the UVLF itself, which can be accommodated by very efficient star formation on top of the steep halo mass function, requiring significantly fainter intrinsic observations to discover $z > 16$ galaxies in reasonable volumes. It is unsurprising, therefore, that the unique sensitivity of GLIMPSE -- achieved through a combination of lensing magnification and ultra-deep imaging -- has managed to potentially reveal some of the earliest galaxies in the Universe.

Finally, by evaluating various star-formation scenarios, we demonstrate that $M_{\rm UV} \sim -17$ galaxies at $z \sim 16-17$  can potentially grow into the observed UV-bright galaxies seen by \textit{JWST} at $z \sim 10-14$, without necessarily violating the standard cosmological framework. The extreme SFE and burstiness coupled with a steep HMF are both sufficient to explain the observed trends. This connection further emphasizes the idea that faint, yet numerous, high-$z$ galaxies play a crucial role in the stellar mass assembly of galaxies in the early Universe.

\acknowledgements
We thank the anonymous referee for their thorough feedback, which helped to improve this manuscript.
We thank Steven Finkelstein and Volker Bromm for helpful discussions which helped improve the quality of this manuscript. VK acknowledges support from the University of Texas at Austin Cosmic Frontier Center. HA and IC acknowledge support from CNES, focused on the JWST mission, and the Programme National Cosmology and Galaxies (PNCG) of CNRS/INSU with INP and IN2P3, co-funded by CEA and CNES. The BGU lensing group acknowledges support by grant No.~2020750 from the United States-Israel Binational Science Foundation (BSF) and grant No.~2109066 from the United States National Science Foundation (NSF), and by the Israel Science Foundation Grant No.~864/23. JBM acknowledges support through NSF Grants AST-2307354 and AST-2408637. ASL acknowledges support from Knut and Alice Wallenberg Foundation. AA acknowledges support by the Swedish research council Vetenskapsradet (2021-05559). This work has received funding from the Swiss State Secretariat for Education, Research and Innovation (SERI) under contract number MB22.00072, as well as from the Swiss National Science Foundation (SNSF) through project grant 200020\_207349.  This work is based on observations made with the NASA/ESA/CSA \textit{James Webb Space Telescope}. All of the data presented in this article were obtained from the Mikulski Archive for Space Telescopes (MAST) at the Space Telescope Science Institute. The specific observations analyzed can be accessed via \dataset[doi: 10.17909/4byn-fe55]{https://doi.org/10.17909/4byn-fe55}. These observations are associated with program \#3293. Some of the data products presented herein were retrieved from the Dawn \textit{JWST} Archive (DJA). DJA is an initiative of the Cosmic Dawn Center, which is funded by the Danish National Research Foundation under grant DNRF140. P.N. acknowledges support from the Gordon and Betty Moore Foundation and the John Templeton Foundation that fund the Black Hole Initiative (BHI) at Harvard University where she serves as one of the PIs.

\software{BAGPIPES \citep{carnall19}, BEAGLE \citep{chevallard16}, EAZY \citep{brammer08}, GALFIT \citep{peng02}, grizli \citep{grizli}, sep \citep{sep}, SExtractor \citep{sextractor}}

\facilities{\jwst, \hst}

\clearpage
\appendix
\section{Source Photometry}
In this section we present the AB magnitude for each of high-z candidates presented in \autoref{fig:fig_sample}.
\begin{deluxetable*}{cccccccccc}[h]
\tabcolsep=2mm
\tablecaption{\label{tab:tab_phot} Total AB magnitude within D=0.\farcs{2} circular aperture.}
\tablehead{ID & F090W & F115W & F150W & F200W & F277W & F356W & F410M & F444W & F480M}
\startdata
70467 & $>31.90$ & $>31.90$ & $>31.70$& $>31.80$ & $30.61\pm0.19$ & $30.92\pm0.25$ & $>31.10$ & $30.89\pm0.28$ & $>30.10$ \\
72839 & $>31.60$ & $>31.80$ & $>31.70$& $>31.70$ & $30.55\pm0.20$ & $30.86\pm0.26$ & $30.53\pm0.36$ & $30.87\pm0.27$ & $>30.20$ \\
\enddata
\begin{tablenotes}
\end{tablenotes}
\end{deluxetable*}

\bibliographystyle{aasjournal}
\bibliography{refs}

\end{document}